\documentclass{article}

\usepackage{graphicx}  
\usepackage{amsmath}   
\usepackage{amssymb}   

\def\eq#1{{Eq.~(\ref{#1})}}

\def\ee{E_{\rm equi}}
\def\eh{E_{\rm N}}
\newcommand{\LL}{Lanczos-Lovelock}

  \title{Equipartition energy, Noether  energy and boundary term in gravitational action}
  \author{T. Padmanabhan\\
  IUCAA, Pune University Campus,\\
  Ganeshkhind, Pune- 411 007.\\
  {\small {email: nabhan@iucaa.ernet.in}}
  }

  \date{}

\begin{document}
  
  \maketitle
  
  \begin{abstract}
In the study of horizon thermodynamics and  emergent gravity two natural expressions for energy, $E= 2 TS$ (equipartition energy) and $E=TS$ (Noether energy) arise which differ by a factor 2.  I clarify the role of these two expressions in different contexts and show how $E=TS$ is also closely related to the Noether charge arising from the \textit{boundary term} of the Einstein-Hilbert action.
 \end{abstract}

 
The simplest context in which thermodynamics of horizons comes up is in the case of a Schwarzschild black hole in $D=4$ treated as a solution to Einstein's field equations. In this case, it seems natural to take $E=M$, $T=1/(8\pi M)$ and $S=4\pi M^2$ for energy, temperature and entropy respectively. Elementary algebra tells us that these quantities obey two different relations simultaneously: 
\begin{equation}
 dS = \frac{dE}{T} ; \qquad S = \frac{1}{2}\, \frac{E}{T}
\label{one}
\end{equation} 
Let me begin by looking at these two relations a little bit more closely.  

Consider an injection of energy $dE$ into a system kept at \textit{constant} temperature $T$. Then, integrating the relation $dS = (dE/T)$ at constant $T$ will give $S= E/T$ which differs from the second relation in \eq{one} by a crucial factor of two. On the other hand, if one thinks of an energy transfer from outside the horizon to inside and treat $T = T(E)=1/8\pi E$, as a function of $E$, then the integration of $dS = dE/T(E)$  will lead to $S=(1/2) ET$ which is the relation obeyed by the Schwarzschild black hole. 

For clarity of discussion, let me introduce the terminology \textit{Noether energy} for $E_{\rm N} \equiv TS$ and \textit{equipartition energy} $E_{\rm equi} \equiv 2TS$. (The reason for the names will be clear in the sequel.) The above analysis shows that $E_{\rm N}$ arises when $T$ is held constant while $E_{\rm equiv}$ arises when $T$ is treated as a specified function of $E$ for the system. This immediately brings to mind the  analogy with the canonical ensemble (in which $T$ is held constant) and the micro-canonical ensemble (in which $E$ is held constant). This is broadly what happens in the two physical situations in which the energies $\ee$ and $\eh$ occur. It is prudent --- even in classical general relativity ---not to enter into discussions as to what is the ``correct`` expression for energy or which energy is ``more physical`` etc.; it is even more so in the case of emergent gravity and related areas which are still in their infancy. I shall therefore confine myself to describing certain specific results and contexts connected to  $\ee$ and $\eh$  and their inter-relationships.

The expression for $\ee$ in the context of a general horizon was introduced by me in the form $S=(1/2)\beta E$ in 2004 in the case of Einstein gravity \cite{cqgpap}. This expression is equivalent to $\ee = (1/2) n k_BT$ where $n=(A/L_P^2)$ is the number of degrees of freedom in an area $A$ if we attribute one bit per Planck area. This has a compelling naturalness as an equipartition law. More remarkably, the idea extends \cite{surfaceprd} to  all \LL\ models! For a \LL\ model with an entropy tensor $P^{ab}_{cd}\equiv \partial L/\partial R_{ab}^{cd}$ and one can write
\begin{equation}
\ee = \frac{1}{2}k_B\int_{\partial\cal V} dn T_{loc}; \qquad
 \frac{dn}{dA}=\frac{dn}{\sqrt{\sigma}d^{D-2}x}=32\pi P^{ab}_{cd}\epsilon_{ab}\epsilon^{cd}
\label{diffeoeqn}
\end{equation} 
where $\epsilon_{ab}$ is the binormal on the codimension-2 cross-section.
This has a (holographic) interpretation in terms of the  equipartition of the degrees of freedom on an equipotential boundary surface enclosing the energy responsible for gravity (which is the Komar energy in general relativity with a suitable generalization for \LL\ models). Thus, $\ee$ seems to have direct physical relevance in more than one context and, of course, $\ee = Mc^2$ for a Schwarzschild black hole.

The energy $\eh$, on the other hand, arises quite naturally whenever we consider transfer of energy across any null surface as viewed by a local Rindler observer for whom the null surface acts as a local Rindler horizon. I have discussed its role in introducing an observer dependence to thermodynamics in, for example, Section 4.4 of Ref.~\cite{rop}. This expression and its integrated version $S= E/T$ are directly applicable to those contexts in which the injected energy is not considered as a part of a self-gravitating system and we  keep the temperature of the horizon constant in spite of the injection of the energy. In Ref.~\cite{rop} and elsewhere, I have argued that any energy injected onto a null surface appears to hover just outside the horizon for a very long time as far as the local Rindler observer is concerned and thermalizes at the temperature of the horizon if it is assumed to have been held fixed. This is a local version of the well known phenomenon that, the energy dropped into a Schwarzschild black hole horizon hovers just outside $R=2M$ as far as an outside observer is concerned. In the case of a local Rindler frame, similar effects will occur as long as the Rindler acceleration is sufficiently high; that is, if $\dot\kappa/\kappa^2 \ll 1$. Obviously, the physical role $\ee$ is quite different from that of $\eh$. (As an aside, let me mention that the same factor two difference arises in a different context; see \cite{visser})

There is another  context in which $\eh$ appears in a  natural manner which has the advantage that the ideas can be extended to a much wider class of gravitational theories. It is well known that the expression for the Wald entropy of a horizon can be written as 
\begin{equation}
S_{\rm Wald} \equiv   \frac{1}{T}  \int d^{D-2}\Sigma_{ab}\, J^{ab}
\label{noetherint}
\end{equation}
where $J^{ab}$ is the Noether potential corresponding to the Killing vector and the integral is over the horizon slice.
This immediately tells us that 
\begin{equation}
 \eh = TS_{\rm Wald} =  \int d^{D-2}\Sigma_{ab}\, J^{ab}
\end{equation} 
is\textit{ nothing but the Noether charge}! (This is sufficient reason to call $\eh=TS$ as the Noether energy.) In the case of a null surface acting as a local Rindler horizon, one 
can define $\eh$ in terms of the Noether charge corresponding to the approximate 
 local Killing vector. The interpretation of the  Noether charge as an energy rather than an entropy  has been noted previously in the literature, especially by Hayward (see e.g., eq 61 and the discussion around it in \cite{hayward}). In this form, the expression for $\eh$ immediately generalizes to \LL\ models of gravity which is rather gratifying since I believe any correct result related to horizon thermodynamics must transcend general relativity and hold true for all \LL\ models.

The same expression (and interpretation) also arises in the interpretation of gravitational field equations as an entropy balance law (see, e.g \cite{entdenspacetime}) $\delta S_m=\delta S_{\rm grav}$  across the null surface which works for all \LL\ models. The expressions actually equated are
$
\beta  T^{aj}\xi_a \xi_j\ dV_{prop}  
$
and
$
\beta \xi_a J^a \ dV_{prop}\to 2 \beta \mathcal{G}^{aj}\xi_a \xi_j \ dV_{prop}
$
at the appropriate limit, where $\mathcal{G}^{aj}$ is the analogue of the Einstein tensor in the \LL\ models. The factor $\beta=2\pi/\kappa$ cancels out  in this expression --- as it should, since the local Rindler observer with a specific $\kappa$ was introduced only for interpretational convenience --- and one can also work with the relation $T\delta S_m=T\delta S_{\rm grav}$ where the right hand side is essentially the change in Noether energy $\delta\eh$ of the horizon. I stress that \textit{all this is possible for a general \LL\ model}.

The physical context in which $\eh$ plays a role requires the local Rindler horizon to be treated as  a  system (like a hot metal plate) at a given temperature and possessing certain intrinsic degrees of freedom. Such a context has been explored in the literature by many people, notably by Carlip \cite{carlip} who has argued that the universality of horizon entropy arises because only horizon degrees of freedom plays a crucial role. Recently, we have been able to relate Carlip's program based on Virasoro algebra with the Noether current approach and generalize the results to all \LL\ models \cite{TPbm1}.  

More specifically, we could show \cite{TPbm2} that: (a) There is a close relation between this approach and the \textit{boundary term of the gravitational action} and (b) the local Rindler observer who perceives the null surface as a horizon will attribute to it certain (effective) degrees of freedom which are not recognized by, say, a freely falling observer who sees no thermal effects.
Given the fact that $\eh=TS$ and the Noether charge  are one and the same, it is obvious that there should exist a simple relationship between the boundary term in the gravitational action and $\eh$. This is indeed the case and I will briefly describe how it comes about. 

The Gibbons-Hawking-York surface term in general relativity is given by
\begin{eqnarray}
\mathcal{A}_{sur} = \frac{1}{8\pi }\int_{\partial\mathcal{V}} \sqrt{h}d^3x K
 =\frac{1}{8\pi }\int_{\mathcal{V}} \sqrt{-g}d^4x\nabla_a(Kn^a)~,
\label{1.34}
\end{eqnarray}
where $n^a$ is the unit normal to the boundary $\partial{\mathcal{V}}$ of the region ${\mathcal{V}}$ and $K=-\nabla_an^a$ is the trace of the extrinsic curvature of this boundary. Since the Lagrangian is a scalar, the Noether current $J^a\equiv \nabla_bJ^{ab}$ for a diffeomorphism $x^a\rightarrow x^a+\xi^a$ can be found by considering the changes of  both sides of \eq{1.34} as the Lie derivative and then equating them (see, e.g., the Appendix of \cite{TPbm2}). The Noether potential $J^{ab}$ is then given by:
\begin{eqnarray}
J^{ab} = \frac{K}{8\pi }\Big(\xi^an^b - \xi^bn^a\Big)~.
\label{1.35}
\end{eqnarray}
An elementary calculation in the local Rindler frame now shows that the $\eh$ is given by
\begin{equation}
 \eh =  \int d^{D-2}\Sigma_{ab}\, J^{ab} = \frac{\kappa A_\perp}{8\pi} =TS
\end{equation}
where $A_\perp$ is the transverse area of the horizon. It is also easy to verify by explicit computation that one gets the same result for all standard black hole horizons, as one should.

In the above analysis we used the Noether current arising from the \textit{boundary term of the action}  in order to stress the conceptual point that the results are closely related to the horizon surface. On the other hand, we know very well that the Noether potential $ J_{ab} = (16\pi)^{-1}[\nabla_a \xi_b - \nabla_b \xi_a]$ corresponding to the full Einstein-Hilbert Lagrangian $ L = R/16\pi $ also leads to the same Noether charge $(\kappa A_\perp/8\pi)$.   So one could have interpreted $\eh$ in terms of either Noether potential but I prefer the interpretation based on surface term in the action.

One motivation for writing this note stems from the recent interest in $\eh=TS$ in a few papers \cite{TS} which do not mention the connection between $\eh$ and the Noether charge, \textit{viz., that they are the same} and $\eh$ is not a physical entity unrelated to previously known expressions! The relationship between $\eh$ and the boundary term of the gravitational action (which is essentially the relationship between the Noether charge and the boundary term of the action, a relationship that is probably of deeper significance) also seems to have gone unnoticed earlier. While  this note was in the final stages of preparation, two papers appeared in the arXiv \cite{lee} which related $\eh$ to spinfoam based models and their boundary action, etc. However, as pointed out above, the relationship is actually very simple. It holds for the standard general relativistic action and its boundary term and is physically transparent once the connection between the Noether charge and $\eh$ is recognised. 

I thank B.R.Majhi for useful discussions.


 \end{document}